\documentclass[cits]{PoS}
\newcommand{\preprint}{\newline%
  \begin{picture}(0,0)
  \put(240,107){\rm\small RUG-CTN 358, DESY 08-137, HU-EP-08/49}
  \end{picture}}

\title{Status of ETMC simulations with $N_{\rm f}=2+1+1$ twisted mass fermions\preprint}

\ShortTitle{Status of ETMC simulations with $N_{\rm f}=2+1+1$ twisted mass fermions}

\author{R\'emi Baron$^a$, Philippe Boucaud$^{b}$, Albert Deuzeman$^{c}$, Vincent Drach$^{d}$, \mbox{Federico Farchioni}$^{e}$, Vicent Gimenez$^{f}$, Gregorio Herdoiza$^g$, Karl Jansen$^g$, \mbox{Istv\'an Montvay}$^{h}$, David Palao$^{f}$, Elisabetta Pallante$^{c}$, Olivier P\`ene$^{b}$, \speaker{Siebren Reker}\thanks{For the ETM Collaboration} $^{c}$\footnote{Email: {s.f.reker@rug.nl}}, Enno E. Scholz$^{i}$, Carsten Urbach$^j$, Marc Wagner$^j$ and Urs Wenger$^k$\\
\\\llap{$^a$}CEA, Centre de Saclay, IRFU/Service de Physique Nucl\'eaire, F-91191 Gif-sur-Yvette, France\\
\llap{$^b$}Laboratoire de Physique Th\'eorique (B\^at. 210), Universit\'e de Paris XI, Centre d'Orsay, 91405 Orsay-Cedex, France\\
\llap{$^c$}Centre for Theoretical Physics, University of Groningen, Nijenborgh 4, 9747 AG Groningen, the Netherlands\\
\llap{$^d$}Laboratoire de Physique Subatomique et Cosmologie, 53 avenue des Martyrs, 38026 Grenoble, France\\
\llap{$^e$}Universit\"at M\"unster, Institut f\"ur Theoretische Physik, Wilhelm-Klemm-Stra\ss e 9, D-48149 M\"unster, Germany\\
\llap{$^f$}Dep. de F\'isica Te\`orica and IFIC, Universitat de Val\`encia-CSIC, Dr.Moliner 50, E-46100 Burjassot, Spain\\
\llap{$^g$}DESY, Platanenallee 6, D-15738 Zeuthen, Germany\\ 
\llap{$^h$}Deutsches Elektronen-Synchrotron DESY, Notkestr. 85, D-22603 Hamburg, Germany\\
\llap{$^i$}Physics Department, Brookhaven National Laboratory, Upton, NY 11973, USA\\
\llap{$^j$}Humboldt-Universit\"at zu Berlin, Institut f\"ur Physik, Newtonstra\ss e 15, D-12489 Berlin, Germany\\
\llap{$^k$}Institute for Theoretical Physics, University of Bern, Sidlerstr. 5, CH-3012 Bern, Switzerland\\}

\abstract{We present the status of runs performed in the twisted mass formalism with $N_{\rm f}=2+1+1$ flavours of dynamical fermions: a degenerate light doublet and a mass split heavy doublet. The procedure for tuning to maximal twist will be described as well as the current status of the runs using both thin and stout links. Preliminary results for a few observables obtained on ensembles at maximal twist will be given. Finally, a reweighting procedure to tune to maximal twist will be described.}

\FullConference{The XXVI International Symposium on Lattice Field Theory\\July 14-19 2008\\Williamsburg, Virginia, USA}

\begin{document}

\section{Introduction}
The twisted mass formulation of Lattice QCD (see for example references \cite{Frezzotti:2000nk,TM intro}) is being studied extensively with $N_{\rm f}=2$ dynamical flavours by the European Twisted Mass (ETM) collaboration \cite{Boucaud:2007uk, Boucaud:2008xu,Urbach:2007rt,Alexandrou:2007qq,Alexandrou:2008tn}. In the twisted mass formulation of QCD, the Wilson term is chirally rotated within a doublet. For the $2$-flavour case, this amounts to a mass degenerate $(u,d)$ quark doublet. In recent studies, a partially quenched setup \cite{Blossier:2007vv} was used to add the strange quark to this formulation. One way to include a dynamical strange quark is to add in addition to the strange quark a charm quark in a heavier and mass-split doublet. This formulation was introduced in \cite{Frezzotti} and first explored in \cite{Chiarappa}. We will use that formulation and report on the current status of runs with $N_{\rm f}=2+1+1$ twisted mass quarks conducted by the ETM collaboration. In section \ref{secaction} we will detail the lattice action and the simulation algorithm, while in section \ref{sectuning} we describe our procedure for tuning to maximal twist. In section \ref{subsecreweight} we discuss a reweighting procedure allowing to speed up and improve the accuracy of tuning to maximal twist. In section \ref{secresults} we show a set of preliminary results. Finally in section \ref{secstout} we describe first results from one level of stout smearing \cite{Stout}.
\section{Lattice action and simulation algorithm}\label{secaction}
In the gauge sector we use the Iwasaki \cite{Iwasaki:1985we} gauge action (with $\beta=1.9$ in the ensembles we are discussing here). With this gauge action we observe a smooth dependence of phase sensitive quantities on the hopping parameter $\kappa$ around its critical value ($\kappa_{\rm crit}$). The fermionic action for the light doublet is given by:
\begin{equation}
S_{l}=a^4\sum_x  \left\{\bar{\chi_l}(x)\left[D_W[U] + m_{0,l} + i \mu_l \gamma_5\tau_3  \right] \chi_l(x)\right\}
\end{equation}
\cite{TM intro}, where $m_{0,l}$ is the untwisted bare quark mass tuned to its critical value $m_{\rm crit}$, $\mu_{l}$ is the bare twisted quark mass for the light doublet, $\tau_3$ is the third Pauli matrix acting in flavour space and $D_{W}[U]$ is the Wilson-Dirac operator. The quark doublet fields that are proportional to the renormalized $u$ and $d$ doublet fields in the physical basis are given by:
\begin{equation}
\psi_{l}^{phys} = e^{\frac{i}{2}\omega_{l}\gamma_{5}\tau_{3}}\chi_{l}, \qquad \bar{\psi}_{l}^{phys}=\bar{\chi}_{l}e^{\frac{i}{2}\omega_{l}\gamma_{5}\tau_{3}}.
\end{equation}
The light doublet twisting angle $\omega_{l}$ takes the value $\omega_{l} \to \pm \frac{\pi}{2}$ as $m_{0,l} - m_{\rm crit} \to \pm 0$. In the heavy sector, as has been shown by Frezzotti and Rossi in \cite{Frezzotti}, a \emph{real quark determinant} with a mass-split doublet can be obtained if the mass splitting is taken to be orthogonal in isospin space to the twist direction. The mass term in the heavy sector then becomes:
\begin{equation}
S_{h}=a^4\sum_{x}\left\{\bar{\chi}_{h}(x)\left[m_{0,h}+i\mu_{\sigma}\gamma_{5}\tau_{1}+\mu_{\delta}\tau_{3}\right]\chi_{h}(x)\right\}.
\end{equation}
This time the mass splitting is in the $\tau_{3}$ direction and the twist in the $\tau_{1}$ direction, which yields a similar rotation relation to the physical quark basis:
\begin{equation}
\psi_{h}^{phys} = e^{\frac{i}{2}\omega_{h}\gamma_{5}\tau_{1}}\chi_{h}, \qquad \bar{\psi}_{h}^{phys}=\bar{\chi}_{h}e^{\frac{i}{2}\omega_{h}\gamma_{5}\tau_{1}}.
\end{equation}
At maximal twist ($|\omega_{l}|=|\omega_{h}|=\frac{\pi}{2}$) physical observables are automatically $\cal{O}$$(a)$ improved without the need to determine any action or operator specific improvement coefficients. The gauge configurations are generated with two different algorithms, depending on whether thin or stout links are being used. For thin links a Polynomial Hybrid Monte Carlo (PHMC) updating algorithm \cite{Chiarappa:2005mx,Urbach:2005ji}, while for stout links the PHMC setup described in \cite{stoutalgo} is used.\footnote{Note that in the Monte Carlo simulation we monitor that the minimal eigenvalue of the Dirac operator of the heavy doublet always remains positive.} All ensembles described in this paper correspond to lattices with spatial extension $L/a=24$ and temporal extension $T=2L$. Our lightest twisted mass quark value ($a\mu_{l}=0.004$) corresponds to a pion mass of around $300$ MeV and a value of $m_{\pi}L$ around $3.5$.
\section{Tuning action parameters}\label{sectuning}
\subsection{Procedure}\label{subsectunproc}
Tuning to maximal twist requires to set $m_{0,l}$ and $m_{0,h}$ equal to some proper estimate of the critical mass $m_{\rm crit} = m_{\rm crit}(\beta)$ \cite{Frezzotti}. Here we set $m_{0,l}=m_{0,h} \equiv 1/(2\kappa)-4$ and, at fixed $\mu_{\sigma}$ and $\mu_{\delta}$, for each value of $\mu_{l}$, we tune $\kappa$ so as to obtain $am_{\rm PCAC,l}=0$ (see also eq. (\ref{eqpcacmass})). Note that the untwisted PCAC mass arising from the heavy sector is $\cal{O}$$(a)$ and $\omega_h$ is automatically very close to $\frac{\pi}{2}$ \cite{Chiarappa}. Our way of implementing maximal twist can be viewed as the extension to the theory with $2+1+1$ dynamical quarks of the optimal critical mass method of Refs.~\cite{MTNF2}. In this way we expect \cite{NOTE} the (absolute) magnitude of lattice artifacts to remain roughly constant as $\mu_l \to 0$. The PCAC mass in the light sector is defined in the following way:
\begin{equation}\label{eqpcacmass}
m_{\rm PCAC,l}=\frac{\sum_{x}\left< \partial_{0}A^{a}_{0,l}(x,t)P_{l}^{a}(0)\right>}{2\sum_{x}\left<P^{a}_{l}(x,t)P^{a}_{l}(0)\right>},
\end{equation}
where the axial and pseudoscalar bilinears $A_{\mu,l}$ and $P_{l}$ are
\begin{equation}
A^{a}_{\mu,l}\equiv\bar{\chi}_{l}\gamma_{\mu}\gamma_{5}\frac{\tau_{a}}{2}\chi_{l} \qquad \textrm{and} \qquad
P^{a}_{l}=\bar{\chi}_{l}\gamma_{5}\frac{\tau_{a}}{2}\chi_{l}.
\end{equation}
The numerical precision at which the condition $m_{\rm PCAC,l}=0$ is fulfilled in order to avoid residual $\cal{O}$$(a^2)$ effects when the pion mass is decreased is, for the present range of lattice spacings, $|\epsilon/\mu_{l}| \lesssim 0.1$, where $\epsilon$ is the small deviation of $m_{\rm PCAC,l}$ from zero \cite{Boucaud:2008xu, Dimopoulos:2007qy}. Note that typically a good statistical precision on the estimate of $m_{\rm PCAC,l}$ is required to satisfy this condition. This computationally disadvantageous situation can be remedied through the use of reweighting, as described in section \ref{subsecreweight}. 

The heavy doublet mass parameters $\mu_\sigma$ and $\mu_\delta$ should be adjusted in order to reproduce the phenomenological values
of the renormalized $s$ and $c$ quark masses. The latter are related to $\mu_\sigma$ and $\mu_\delta$ via \cite{Frezzotti}:
\begin{equation}
(m_{s,c})_{\rm R} = \frac{1}{Z_{P}} (\mu_\sigma \mp \frac{Z_{P}}{Z_{S}} \mu_\delta)
\end{equation}
where the $-$ sign corresponds to the strange and the $+$ sign to the charm. In practice we fix the values $\mu_\sigma$ and $\mu_\delta$ by requiring
the kaon mass to take its physical value and $(m_c)_{\rm R}\gtrsim 10 (m_s)_{\rm R}$ (through an estimate of $Z_P/Z_S$).

\subsection{Status}\label{subsectunstat}
\begin{figure}[h]
\begin{minipage}[ht]{7.5cm}
\includegraphics[width=\textwidth]{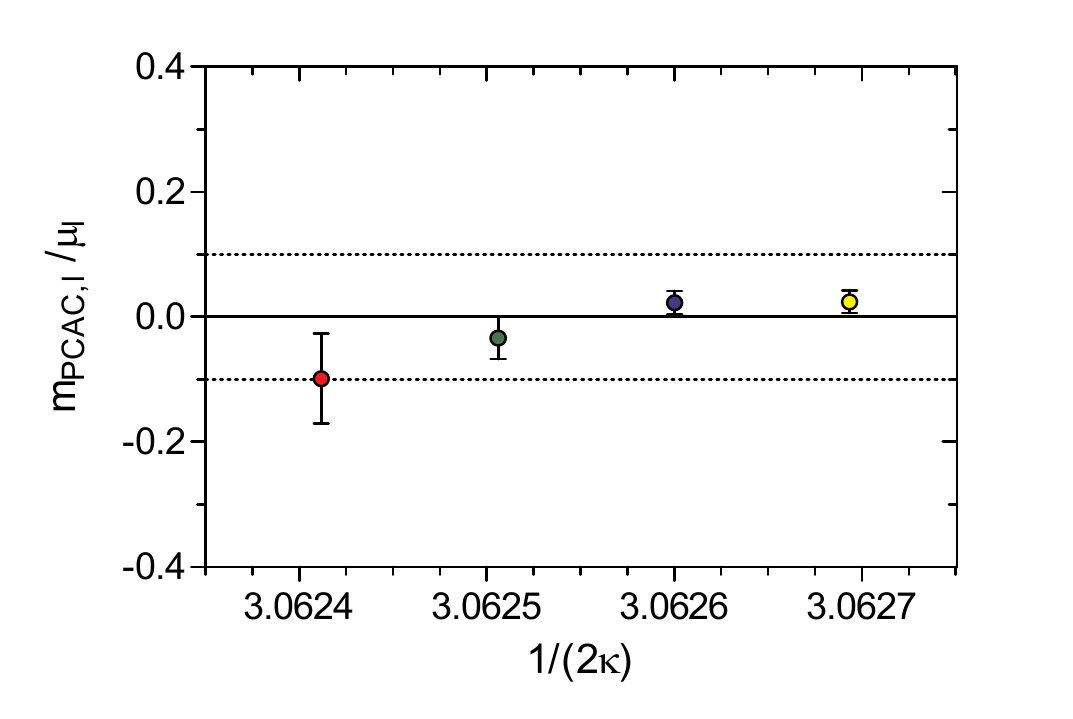} 
\caption{Tuned runs at $\beta=1.9$, with a box size of $24^{3}\times 48$, with $a\mu_{\sigma}=0.15$ and $a\mu_{\delta}=0.19$. From left to right the points correspond to $a\mu_{l}=0.004$ (red), $a\mu_{l}=0.006$ (green), $a\mu_{l}=0.008$ (blue) and $a\mu_{l}=0.01$ (yellow). For all ensembles, $m_{\rm PCAC,l}/\mu_{l}$ is within the $10\%$ level (see text).}
\label{figtuning}
\end{minipage}
\hspace{0.5cm}
\begin{minipage}[ht]{7.5cm}\vspace*{-0.1cm}
\includegraphics[width=\textwidth]{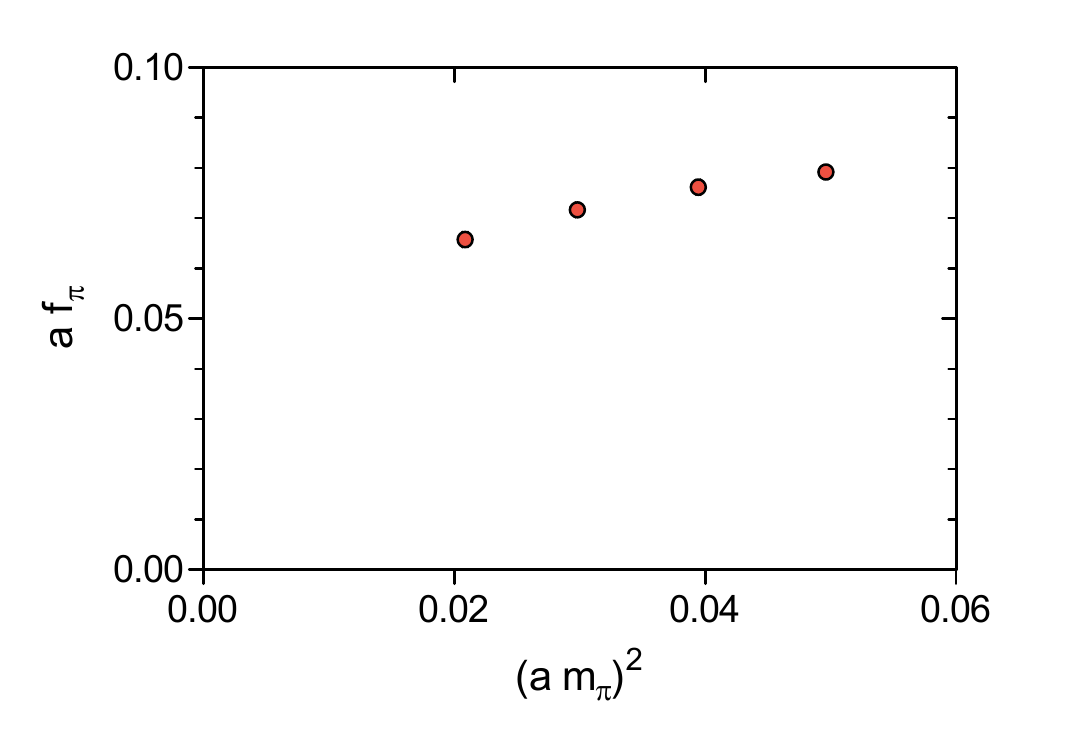} 
\caption{$af_\pi$ as a function of $(am_{\pi})^2$ for the four ensembles described above. All of these are at $\beta=1.9$, $L/a=24$ and $T/a=48$. Error bars are plotted, but they fall within the size of the dots.}
\label{figfpi}
\end{minipage}
\end{figure}

The tuning status of our ensembles is summarized in table \ref{tabruns} and illustrated in figure \ref{figtuning}. The ensembles considered here have the following parameters: $\beta=1.9$, $a\mu_{\sigma}=0.15$ and $a\mu_{\delta}=0.19$, with lattice volumes of $24^{3}\times 48$.
\begin{table}[h]
\begin{center}
\begin{tabular}{|c|c|c|c|c|}
\hline
$a\mu_l$ & $0.004$ & $0.006$ & $0.008$ & $0.01$\\
\hline
$\kappa_{\rm crit}$ & $0.163270$ &  $0.163265$ &  $0.163260$ & $0.163255$\\
\hline
\end{tabular}
\end{center}
\caption{Summary of tuning to maximal twist.}
\label{tabruns}
\end{table}

Tuning to $\kappa_{\rm crit}$ was performed independently for each $\mu_{l}$ value. From table \ref{tabruns} we observe that the estimate of $\kappa_{\rm crit}$ depends weakly on $\mu_{l}$. Figure \ref{figtuning} shows that for all $\mu_{l}$ values the condition $|\epsilon/\mu_{l}|\lesssim0.1$ has been fulfilled.
\subsection{Reweighting}\label{subsecreweight}
Reweighting towards the chiral limit for (untwisted) Wilson fermions has been recently explored in reference \cite{HasenfratzHoffmannSchaefer} (and see also the conference contribution \cite{Luscher:2008tw} with further references). For twisted Wilson fermions one can use reweighting near maximal twist for tuning the last digits of the hopping parameter $\kappa$. Once the update is close to $\kappa_{\rm crit}$ (in the present case for $\Delta\kappa=|\kappa-\kappa_{\rm crit}|={\cal O}(10^{-5})$) the situation is rather advantageous for an effective reweighting because the gauge field configuration is fluctuating between positive and negative values of $m_{\rm PCAC,l}$. Thus, in order to complete the tuning one has to just reweight the ensemble to a close-by value of $\kappa$ at which $m_{\rm PCAC,l}=0$ is realized. 

For calculating the ratio of determinants one could use the methods developed in reference \cite{HasenfratzHoffmannSchaefer}, but we applied a simple and fast method based on least-square optimized polynomial approximations introduced in \cite{TSMB}. In the case of $N_{\rm f}$ flavours, these polynomials satisfy
\begin{equation}
\lim_{n\to\infty} P^{(1)}_n(x) = x^{-N_{\rm f}/2} ,
\hspace{2em}
\lim_{n\to\infty} P^{(2)}_n(x) = \frac{1}{\sqrt{P^{(1)}_{n_1}(x)}} ,
\hspace{2em}
x \in [\epsilon,\lambda]
\end{equation}
where the interval $[\epsilon,\lambda]$ covers the spectrum of $Q^\dagger(\kappa) Q(\kappa)$, with $Q$ the fermion matrix and $n_1$($n_i$) the actual value of $n$ used in the case of the polynomial $P^{(1)}$($P^{(i)}$). With these polynomials the reweighting factor for $\kappa_1\to\kappa_2$ is $1/\det[P^{(2)}_{n_2}(\kappa_1) P^{(1)}_{n_1}(\kappa_2) P^{(2)}_{n_2}(\kappa_1)]$. A stochastic estimator of the reweighting factor can be obtained from a random Gaussian vector as
\begin{equation}
\exp\{-\eta^\dagger(\Omega-1)\eta\} ,
\hspace{1em} {\rm with} \hspace{1em}
\Omega \equiv P^{(2)}_{n_2}(\kappa_1) P^{(1)}_{n_1}(\kappa_2)
              P^{(2)}_{n_2}(\kappa_1) \ .
\end{equation}
We showed in test runs that with a few stochastic estimators an effective reweighting can be achieved for $\Delta\kappa\simeq 10^{-5}$. We actually applied {\it determinant break-up} by a factor of 2, that is, we considered two copies of half doublets, both for light and heavy doublets. With this method the number of long tuning runs for maximal twist can be reduced.
\section{Results}\label{secresults}
\begin{figure}[h]
\begin{minipage}[ht]{7.5cm}
\includegraphics[width=\textwidth]{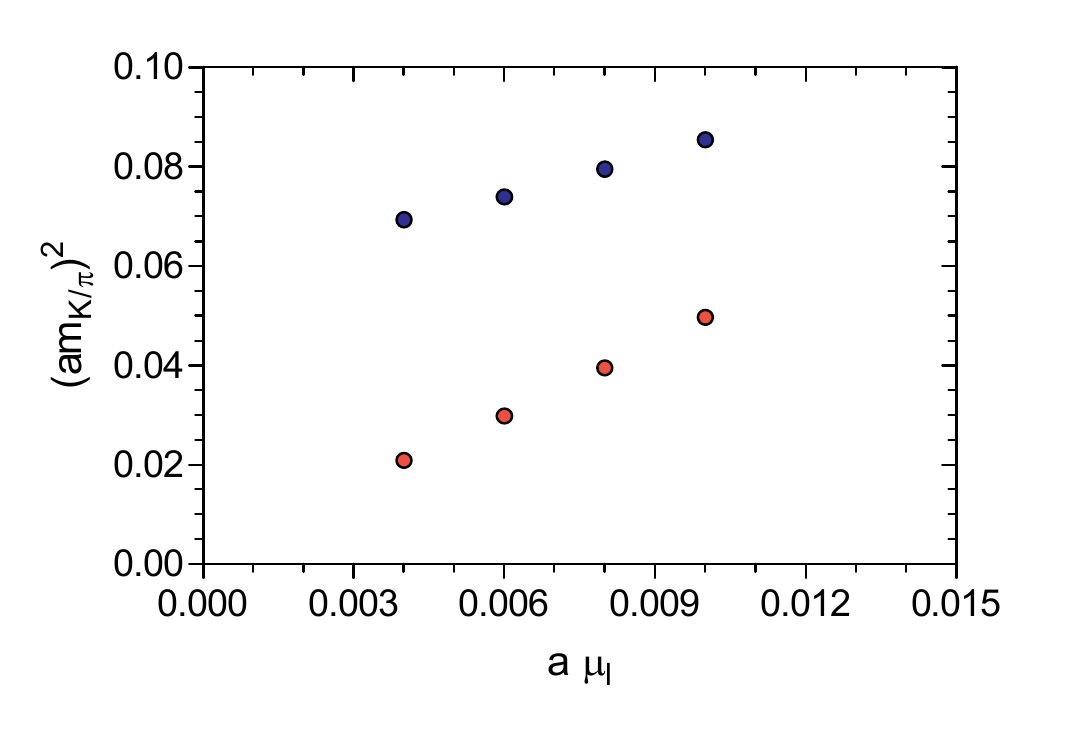} 
\caption{The kaon (blue) and pion (red) squared lattice masses are plotted versus the bare quark mass $a\mu_{l}$.}
\label{figkaon}
\end{minipage}
\hspace{0.5cm}
\begin{minipage}[ht]{7.5cm}\vspace*{-0.1cm}
\includegraphics[width=\textwidth]{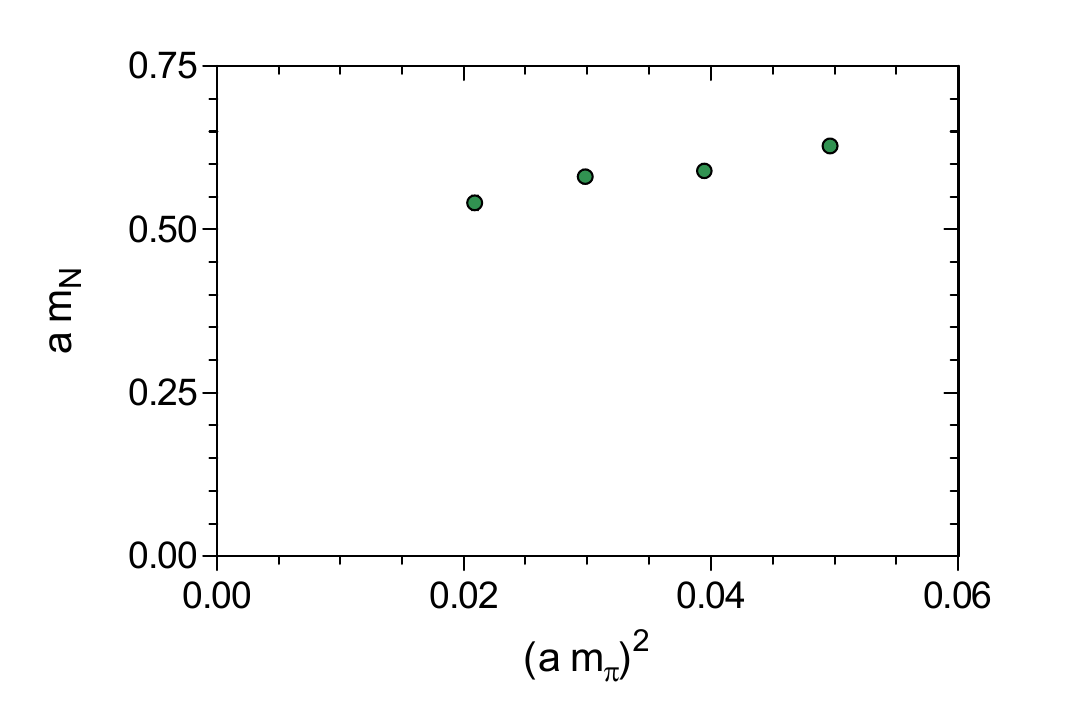} 
\caption{The nucleon mass ($am_{\rm N}$) as a function of $(am_{\pi})^2$ for the four ensembles described above. A subset of the available statistics has been used here.}
\label{fignucleon}
\end{minipage}
\end{figure}

In this section we present preliminary results for the light quark mass dependence of the pion decay constant and the pion, kaon and nucleon masses (figures \ref{figfpi},\ref{figkaon} and \ref{fignucleon}). The pion masses in these ensembles are in the range from $300$ to $500$ MeV, corresponding to $\beta=1.9$ and $a\mu_{l}=0.004$ to $0.01$ (see section \ref{subsectunstat}).

We measured $r_{0}/a$ and performed its chiral extrapolation. The results are compared to those of our $N_{\rm f}=2$ runs with $\beta=3.9$ at the same lattice volume in table \ref{tabcomp}. The vicinity of the two values is very encouraging, since it might allow to perform a direct comparison between these two ensembles and identify possible effects due to the dynamical strange and charm quarks.
\begin{table}[h]
\begin{center}
\begin{tabular}{|c|c|c|c|c|c|c|}
\hline
& $S_{\textrm{Gauge}}$ & $\beta$ & $r_{0}/a$  in chiral limit  \\
\hline
$N_{\rm f}=2$ & tl-Sym & $3.9$ & $5.22(2)$ \\ 
$N_{\rm f}=2+1+1$ & Iwasaki & $1.9$ & $5.18(5)$\\
\hline
\end{tabular}
\caption{Comparison of $r_0/a$ in the chiral limit between the $N_{\rm f}=2$ and $N_{\rm f}=2+1+1$ runs }
\label{tabcomp}
\end{center}
\end{table}
\section{Stout smearing}\label{secstout}
We separately tested the effectiveness and impact on observables of one level of stout smearing \cite{Stout} with $\rho=0.15$ (in every direction). In order to perform this comparison, we tuned a run with stout smearing at $a\mu_{l}=0.004$ to maximal twist. The parameters of this run are compared with the one with thin links in \mbox{table \ref{tabstout}}.
\begin{table}[h]
\begin{center}
\begin{tabular}{|c|c|c|c|c|c|c|}
\hline
$\rho$ & traj & $\kappa_{\rm crit}$ & $a\mu_{l}$ & $a\mu_{\sigma}$ & $a\mu_{\delta}$\\
\hline
$0$ & $\sim5000$ & $0.16327$ & $0.004$ & $0.15$ & $0.19$ \\ 
$0.15$ & $\sim2500$ & $0.14552$ & $0.004$ & $0.17$ & $0.185$ \\
\hline
\end{tabular}
\caption{Comparison of the ensembles with $a\mu_{l}=0.004$, with ($\rho=0.15$) and without ($\rho=0$) stout smearing.}
\label{tabstout}
\end{center}
\end{table}
For the simulation with stout smearing the values of the heavy doublet mass parameters $\mu_\sigma$ and $\mu_\delta$ were chosen according to
the same criterion as described in section \ref{subsectunproc}. The difference in the values for $\mu_\sigma$ and $\mu_\delta$ reported in Table \ref{tabstout} is likely due to the fact that the ratio of the renormalization factors $Z_P/Z_S$ increases when replacing thin links with stout links. Although the statistics are somewhat different between the two runs we can still make a few conclusive comparisons. We observe that our estimate of $\kappa_{\rm crit}$ decreases with stout links. In table \ref{tabstout2} we show a weak dependence of $r_{0}/a$ on $a\mu_{l}$ for simulations with stout links. Finally we observe that at fixed bare quark mass $a\mu_{l}$ the pion mass (in physical units) from stout link simulations decreases with respect to that from thin link simulations. This finding can be ascribed to an increase of the value of the renormalization constant $Z_P$ (implying a decrease of the renormalized light quark mass).
\begin{table}[h]
\begin{center}
\begin{tabular}{|c|c|c|c|c|c|c|}
\hline
$a\mu_l$ & $0.004$ & $0.006$ & $0.008$\\ 
\hline
$r_{0}/a$ &  $4.97(8)$ & $4.90(8)$ & $ 4.99(8)$\\
\hline
\end{tabular}
\caption{Comparison of $r_{0}/a$ values for various ensembles with stout smearing at different light twisted quark mass value $a\mu_{l}$.}
\label{tabstout2}
\end{center}
\end{table}

\section{Conclusions}\label{secconclusions}
We have presented preliminary results for a set of simple observables obtained from ensembles with $N_{\rm f}=2+1+1$ dynamical flavours of maximally twisted Wilson fermions. We have tuned to maximal twist at four values of $\mu_{l}$ on $24^{3}\times 48$ lattices at $\beta=1.9$. A detailed analysis of the data as well as a systematic control of finite size and discretization effects is in progress.

\section*{Acknowledgments}
We thank Roberto Frezzotti for useful discussions and comments on this write-up. We would also like to thank the computer centres of Barcelona, Groningen, Paris (IDRIS), J\"ulich, Munich and Rome (apeNEXT) for providing us with the necessary technical help and computer resources to enable us to perform this work.

\end{document}